\documentclass[conference]{IEEEtran}
\IEEEoverridecommandlockouts

\usepackage[utf8]{inputenc} 
\usepackage[T1]{fontenc}    
\usepackage[colorlinks=true, linkcolor=red, citecolor=green, urlcolor=cyan, filecolor=magenta]{hyperref}       
\usepackage{url}            
\usepackage{booktabs}       
\usepackage{amsfonts}       
\usepackage{nicefrac}       
\usepackage{microtype}      
\usepackage{amsmath}
\usepackage{graphicx}
\usepackage{xcolor}
\usepackage{colortbl}
\usepackage{multirow}
\usepackage{float}
\usepackage{subcaption}
\usepackage{enumitem}
\setlist[itemize]{leftmargin=*}
\setlist[enumerate]{leftmargin=*}
\usepackage{array} 
\newcolumntype{P}[1]{>{\centering\arraybackslash}p{#1}}
\usepackage{pifont}

\newcommand{\bi}{\begin{itemize}}
\newcommand{\ei}{\end{itemize}}
\newcommand{\be}{\begin{enumerate}}
\newcommand{\ee}{\end{enumerate}}

\usepackage{makecell}
\usepackage[linesnumbered,ruled,vlined]{algorithm2e}
\usepackage{algorithmic}
\usepackage[skins]{tcolorbox}

\usepackage{amsthm}

\usepackage{wrapfig}

\usepackage[skins]{tcolorbox}

\newenvironment{RQ}{\vspace{2mm}\begin{tcolorbox}[enhanced,width=\linewidth,size=fbox,colback=blue!5,drop shadow southeast,sharp corners]}{\end{tcolorbox}}
\usepackage{diagbox}

\begin{document}

\title{Agile Story Point Estimation with Large Language Models}

\author{\IEEEauthorblockN{1\textsuperscript{st} Pranam Prakash Shetty}
\IEEEauthorblockA{\textit{Department of Software Engineering} \\
\textit{Rochester Institute of Technology}\\
Rochester, USA \\
ps9960@rit.edu}
\and
\IEEEauthorblockN{2\textsuperscript{nd} Adarsh Balakrishnan}
\IEEEauthorblockA{\textit{Department of Software Engineering} \\
\textit{Rochester Institute of Technology}\\
Rochester, USA \\
ab3562@rit.edu}
\and
\IEEEauthorblockN{3\textsuperscript{rd} Mengqiao Xu}
\IEEEauthorblockA{\textit{Department of Software Engineering} \\
\textit{Rochester Institute of Technology}\\
Rochester, USA \\
mx7944@rit.edu}
\and
\IEEEauthorblockN{4\textsuperscript{th} Xiaoyin Xi}
\IEEEauthorblockA{\textit{Department of Software Engineering} \\
\textit{Rochester Institute of Technology}\\
Rochester, USA \\
xx4455@rit.edu}
\and
\IEEEauthorblockN{5\textsuperscript{th} Zhe Yu}
\IEEEauthorblockA{\textit{Department of Software Engineering} \\
\textit{Rochester Institute of Technology}\\
Rochester, USA \\
zxyvse@rit.edu}
}

\maketitle

\begin{abstract}
This study investigates the use of large language models (LLMs) for story point estimation. Story points are unitless, project-specific effort estimates that help developers on the scrum team forecast which product backlog items they plan to complete in a sprint. To facilitate this process, machine learning models, especially deep neural networks, have been applied to predict the story points based on the title and description of each item. However, such machine learning models require sufficient amounts of training data (with ground truth story points annotated by human developers) from the same software project to achieve decent prediction performance. This motivated us to explore whether LLMs are capable of (RQ1) predicting story points without training data or (RQ2) with only a few training data points. Our empirical results with four LLMs on 16 software projects show that, without any training data (zero-shot prompting), LLMs can predict story points better than supervised deep learning models trained on $80\%$ of the data. The prediction performance of LLMs can be further improved with a few training examples (few-shot prompting). In addition, a recent study explored the use of comparative judgments (between a given pair of items which one requires more effort to implement) instead of directly annotating the story points to reduce the cognitive burden on developers. Therefore, this study also explores (RQ3) whether comparative judgments are easier to predict than story points for LLMs and (RQ4) whether comparative judgments can serve as few-shot examples for LLMs to improve their predictions of story points. Empirical results suggest that it is not easier for LLMs to predict comparative judgments than to directly estimate the story points, but comparative judgments can serve as few-shot examples to improve the LLMs' prediction performance as well as the human-annotated story points. This study shows that applying LLMs to story point estimation is promising. Good performance can be achieved without labeled data, whereas better performance can be achieved with only a few examples. These examples can also be comparative judgments that require less human effort. 
\end{abstract}

\begin{IEEEkeywords}
Requirement engineering, agile development, story point estimation, comparative judgments, machine learning, large language models
\end{IEEEkeywords}

\section{Introduction}\label{sect1}

Effort estimation plays a critical role in agile software development, directly influencing sprint planning, resource allocation, and delivery predictability. Among various estimation techniques, story points are widely adopted in Scrum-based processes as a relative and unitless measure of development effort \cite{schwaber2020scrum,cohn2005agile}. Story points are typically assigned by developers based on issue titles, descriptions, and contextual knowledge of the project, often through collaborative practices such as Planning Poker \cite{cohn2005agile}. Although effective in practice, this process is inherently subjective, time-consuming, and difficult to scale, particularly for large teams or projects with limited estimation expertise.

To address these challenges, prior research has explored automated story point estimation using machine learning models. Early studies applied traditional regression and shallow learning approaches, whereas later studies demonstrated that deep neural networks could effectively leverage semantic information embedded in backlog item texts. Choetkiertikul et al. showed that deep learning models trained on issue titles and descriptions can outperform classical baselines in predicting story points \cite{choetkiertikul2018deep}. Recently, transformer-based models have been introduced to further enhance estimation accuracy. Fu and Tantithamthavorn proposed GPT2SP, which applies pretrained language models fine-tuned on project-specific data, achieving state-of-the-art performance on multiple datasets \cite{fu2022gpt2sp}.

Despite their success, supervised learning approaches to story point estimation face a fundamental limitation: they require a substantial amount of labeled training data from the same software project. In real-world settings, such data are often scarce, costly to obtain, or unavailable for new or rapidly evolving projects. Additionally, prior work has shown that software effort estimation models often suffer from limited generalizability across projects because of differences in domains and development practices \cite{shepperd2012systematic}. Consequently, the applicability of these models is constrained, particularly in cold-start scenarios.

The emergence of large language models (LLMs) presents new opportunities to address these limitations. LLMs trained on massive corpora have demonstrated strong zero- and few-shot capabilities across a wide range of natural language tasks \cite{brown2020gpt3,wei2022chain}. In software engineering, recent studies have reported that LLMs can support tasks such as code generation, program repair, defect prediction, and requirements engineering \cite{zhang2023llm4se,zheng2023understanding}. Unlike traditional supervised models, off-the-shelf LLMs can perform inference without task-specific training, raising the question of whether they can effectively estimate story points directly from backlog item descriptions. 

In parallel, a recent study by Khan et al. ~\cite{khan2025efficientstorypointestimation} explored comparative judgments as an alternative form of supervision for story point estimation. Instead of assigning absolute story point values, developers compare pairs of backlog items and decide which requires more effort. Grounded in Thurstone’s law of comparative judgment \cite{thurstone1927law} and preference learning studies such as the Bradley--Terry model \cite{bradley1952rank}, comparative judgments have been shown to offer the advantage of reduced cognitive burden and labeling effort. Khan et al. ~\cite{khan2025efficientstorypointestimation} also demonstrated that machine learning models trained on comparative judgments can achieve competitive (sometimes even better) performance while requiring significantly less human annotation effort in story point estimation. Inspired by their work, we explore whether comparative judgments are also easier for LLMs to predict or whether they can serve as effective few-shot examples for guiding LLM-based story point estimation.

In this study, we conducted a systematic empirical study to investigate the use of LLMs for agile story point estimation. Using data from 16 real-world software projects, we examine four research questions: \textbf{RQ1:} How well can LLMs predict story points in a zero-shot setting without any labeled training data? \textbf{RQ2:} Does few-shot prompting with a small number of annotated examples improve prediction performance? \textbf{RQ3:} Do LLMs find comparative judgments easier to predict than absolute story point values? \textbf{RQ4:} Can comparative judgments serve as effective few-shot supervision for improving story point estimation?

Our results show that LLMs can achieve performance comparable to state-of-the-art deep learning models even without training data and that their predictions improve further with minimal supervision. While comparative judgments are not inherently easier for LLMs to predict than story points, they can effectively serve as few-shot examples to enhance estimation accuracy. These findings suggest that LLMs offer a promising, low-cost alternative for agile effort estimation, particularly in data-scarce scenarios, and highlight new opportunities for integrating foundation models into software engineering practices. The code and data used in this study is publicly available at GitHub\footnote{\url{https://github.com/hil-se/LLM_Story_Point}} for reproduction.

\section{Background and Related Work}

\subsection{Story point estimation}

Story points are a relative and unitless measure of development effort widely used in agile software development to support sprint planning and backlog prioritization~\cite{schwaber2020scrum}. Unlike absolute effort measures, such as person-hours, story points capture multiple dimensions of work, including complexity, uncertainty, and implementation effort~\cite{cohn2005agile}. Agile teams often estimate story points collaboratively using structured consensus-based techniques, such as Planning Poker, where team members discuss backlog items and converge toward a shared estimate~\cite{cohn2005agile}. Related expert-based estimation approaches include Wideband Delphi, which relies on iterative anonymous estimation rounds and discussions to reduce bias and improve agreement among estimators~\cite{boehm2000software}.

Although expert-based estimation approaches are widely adopted, they are inherently subjective and can be inconsistent across teams and projects. In addition, estimation accuracy may be influenced by factors such as team experience, incomplete requirements, and cognitive biases~\cite{jorgensen2004review}. These limitations have motivated research on more systematic estimation frameworks. For example, Function Point Analysis (FPA) estimates effort by quantifying software functionality delivered to the user and has been widely studied as a traditional method for cost and effort estimation~\cite{albrecht1979measuring}. Similarly, Use Case Points (UCP) estimate effort based on use case complexity and technical/environmental factors~\cite{karner1993usecase}. While these approaches offer more structured estimation processes, they require careful calibration and may not align well with agile environments in which requirements evolve rapidly.

To reduce the cost and variability of manual story point estimation, prior research has explored automated estimations using machine learning techniques. Early work applied traditional regression and classification models using handcrafted features extracted from backlog items. With the increasing availability of textual issue data, more recent studies have leveraged deep learning models to learn semantic representations directly from issue titles and descriptions~\cite{choetkiertikul2018deep,fu2022gpt2sp,khan2025efficientstorypointestimation}. Choetkiertikul et al.~\cite{choetkiertikul2018deep} introduced a benchmark dataset for story point estimation and demonstrated that deep neural networks can outperform traditional baselines in within-project settings. However, their results also showed a notable performance drop in cross-project scenarios, highlighting the project-specific nature of story point assignments. Fu and Tantithamthavorn~\cite{fu2022gpt2sp} proposed GPT2SP, a transformer-based approach that fine-tunes GPT-2 on project-specific issue data and achieves state-of-the-art performance across multiple projects. Despite their strong within-project performance, these supervised approaches share a fundamental limitation: they require a substantial amount of labeled historical data from the same project to achieve reliable results, and their performance may degrade in data-scarce settings \cite{shepperd2012systematic}.

To address the cognitive burden associated with direct story point annotation, recent studies have explored alternative supervision signals based on relative comparisons. Comparative judgments require developers to decide which of two backlog items requires more effort rather than assigning an absolute story point value. According to Thurstone’s law of comparative judgment \cite{thurstone1927law}, such pairwise comparisons are generally easier and more intuitive for humans. Preference learning frameworks, such as the Bradley--Terry model, provide theoretical foundations for modeling such pairwise comparisons \cite{bradley1952rank}. Building on this idea, Khan et al.~\cite{khan2025efficientstorypointestimation} demonstrated that machine learning models trained on comparative judgments can achieve performance comparable to (if not better than) regression models trained on direct story point values, while requiring less human annotation effort and exhibiting improved data efficiency. These findings suggest that relative supervision can be a promising alternative for reducing labeling costs in story point estimation.

In summary, story point estimation has traditionally relied on expert-based approaches, such as Planning Poker and Delphi-style consensus building, whereas recent studies have introduced data-driven methods using deep learning and transformer-based models. However, existing automated approaches remain constrained by the need for project-specific labeled data and the subjectivity of absolute story point scales. These limitations motivate the exploration of approaches that can operate with little or no labeled data, as well as the investigation of whether large language models can leverage both direct and comparative supervision effectively.

\subsection{Large language models in software engineering}

Large language models (LLMs) have emerged as a powerful paradigm for software engineering (SE) because they combine large-scale pretraining with in-context learning, enabling task adaptation through prompting rather than task-specific model training. Foundational work on GPT-3 demonstrated strong zero-shot and few-shot capabilities across diverse language tasks, while chain-of-thought (CoT) prompting showed that eliciting intermediate reasoning steps can improve performance on complex reasoning problems \cite{brown2020gpt3,wei2022cot}. These properties are particularly relevant to SE, where many tasks involve understanding natural-language artifacts (e.g., requirements, user stories, issue reports, and commit messages) and mapping them to technical decisions. In contrast to conventional supervised approaches that require project-specific labeled data, off-the-shelf LLMs provide a pretrained prior over both natural language and code, making them attractive in cold-start or data-scarce settings.

Recent surveys and empirical syntheses indicate that LLMs are being applied across a broad range of SE activities, including requirements engineering, code generation, code summarization, test generation, bug fixing, program repair, and developer assistance workflows \cite{hou2024llm4se_slr,zhang2023llm4se_survey,zhang2023llm4se_survey}. At the same time, the evaluation landscape for LLM-based SE systems has evolved substantially. Early benchmarks such as CodeXGLUE and HumanEval were important for measuring code understanding and functional code generation in relatively constrained settings \cite{lu2021codexglue,chen2021codex_humaneval}. However, more recent benchmarks such as RepoBench and SWE-bench emphasize repository-level context, cross-file dependency reasoning, retrieval of relevant code context, and multi-step issue resolution, which more closely reflect real-world software development conditions \cite{liu2023repobench,jimenez2023swebench}. This shift is especially important for tasks like effort estimation, where useful signals may depend less on isolated syntax generation and more on contextual interpretation of requirements, ambiguity, scope, and implementation implications.

Despite their promise, LLMs also introduce important challenges for SE applications. Prior studies report risks related to hallucinated APIs or project artifacts, sensitivity to prompt formulation, and variability in outputs across prompts and settings, which can reduce reliability in practical workflows \cite{hou2024llm4se_slr,zhang2023llm4se_survey,eghbali2024dehallucinator}. In code-centric tasks, grounding techniques that iteratively retrieve project-specific APIs or contextual references have been shown to reduce hallucinations and improve generation quality \cite{eghbali2024dehallucinator}. These observations suggest a broader design principle for LLM-based SE systems: performance improves when the model is provided with relevant contextual anchors rather than relying solely on parametric knowledge. In the context of agile story point estimation, this motivates evaluating LLMs not only in zero-shot settings, but also under few-shot prompting and alternative supervision formats (e.g., comparative judgments) that can help calibrate the model to project-specific estimation scales while preserving the practical advantage of low labeling cost.

\section{Methodology}

\begin{figure*}[btp]
    \centering
    \includegraphics[width=\linewidth]{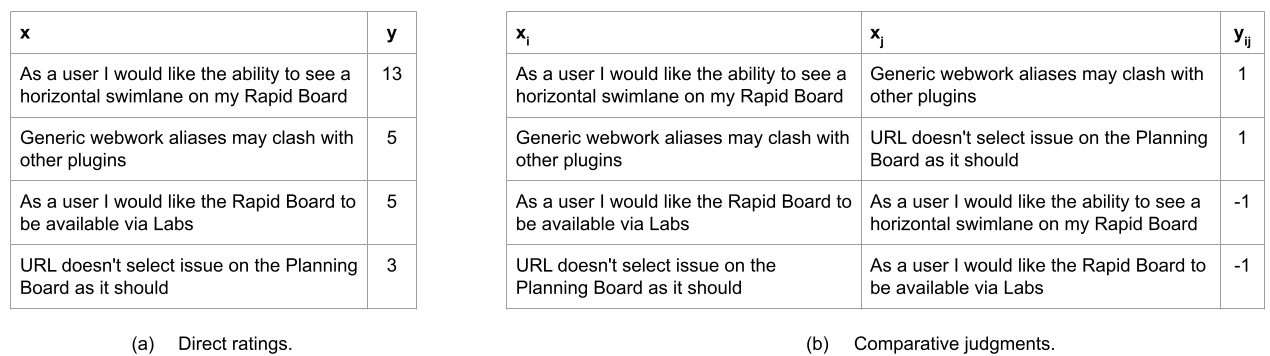}
    \caption{Examples of items with ground truth story points and the generated ground truth comparative judgments.}
    \label{fig:data_gen}
\end{figure*}

This study investigates the use of LLMs in story point estimation and the effectiveness of comparative judgments with the four research questions presented in Section~\ref{sect1}. Experiments were conducted on the story point estimation dataset introduced by Choetkiertikul et al.~\cite{choetkiertikul2018deep}. The dataset consists of backlog item titles and descriptions, as well as their assigned story points, collected through JIRA across 16 projects. For our experiments, we used the same data splits as Fu et al. \cite{fu2022gpt2sp}. The testing split items and their story points are treated as answer sets during evaluation. During the pre-processing step, the titles and descriptions were concatenated to form the item texts used for training and testing~\cite{fu2022gpt2sp}. Example items are shown in Figure~\ref{fig:data_gen}. The ground truth comparative judgments used in RQ3 and RQ4 were generated as $y_{ij} = \text{sgn}(y_i-y_j)$.

To make our conclusions generalizable, we experimented with four off-the-shelf LLMs:
\bi
\item
\textbf{DeepSeek-V3.2}: specifically the \texttt{deepseek-chat} model, is an advanced open-weights reasoning model developed by DeepSeek. It leverages large-scale reinforcement learning (RL) to enhance complex problem-solving abilities and utilizes a Mixture-of-Experts (MoE) architecture for efficient inference, making it highly capable for complex reasoning and coding tasks. With a context window of 128,000 tokens, it supports long-context inputs suitable for extended reasoning tasks. It was accessed via the official DeepSeek SDK. \cite{deepseek2025v32}
\item
\textbf{Gemini Flash Lite}: specifically the \texttt{gemini-2.5-flash-lite} model, is a highly efficient, lightweight model developed by Google, specifically designed for high-speed inference and cost efficiency. With a context window of 1,048,576 tokens, it excels at processing extensive project documents or numerous few-shot examples without truncation. It was accessed via the Google GenAI SDK.\cite{google2025gemini25flashlite}
\item
\textbf{OpenAI GPT-5 Nano}: specifically the \texttt{gpt-5-nano} model, is a compact and highly efficient variant of OpenAI's GPT-5 series, optimized for low-latency inference and cost efficiency while retaining strong instruction-following and reasoning capabilities. With a context window of 400,000 tokens, it supports extensive context processing suitable for long-document tasks. It was accessed via the official OpenAI SDK. \cite{openai2025gpt5nano}
\item
\textbf{Kimi (Moonshot K2)}: specifically the \texttt{moonshotai/kimi-k2-instruct-0905} model, is developed by Moonshot AI and is renowned for its rigorous instruction-following and long-context understanding. With a context window of 262,144 tokens, it excels at retaining extensive contextual histories and performing cross-document reasoning. It was accessed via an OpenAI-compatible inference provider (Groq). \cite{kimi2025k2}
\ei
\subsection{Experimental Setup}
To ensure the reproducibility of our results, all experiments were conducted with the temperature set to 0 to minimize randomness in generation. The maximum number of output tokens was restricted (e.g., 20 tokens) to encourage concise responses suited for extraction. We implemented a robust execution pipeline with automatic retries (up to 3 attempts) to handle potential API instability or rate limits.

\subsection{Response Processing}
Since LLM outputs can vary in format despite prompt instructions, we employed a hierarchical parsing strategy to extract valid story points and comparative decisions:
\bi
\item \textbf{JSON Extraction}: The models were instructed to output data in JSON format (e.g., \texttt{\{"Story point": <integer>\}}). The parsing logic first attempted to decode valid JSON objects from the response.
\item \textbf{Heuristic Fallback}: If JSON parsing failed, we applied regular expression-based heuristics. For story point estimation (RQ1, RQ2, RQ4), we scanned for patterns such as "points: X" or extracted the first distinct integer in the valid range [1, 20]. For comparative judgments (RQ3), we scanned for explicit decision tokens (1 or -1) or key phrases indicating preference.
\item \textbf{Sanitization}: All responses were sanitized to remove Markdown formatting (e.g., bolding like **5**) before extraction to ensure inaccurate parsing did not occur due to formatting artifacts.
\ei

\subsection{RQ1: How do LLMs perform in predicting the story points when there are no training data?}

To explore this research question, we applied the following zero-shot prompt to the four LLMs:

\begin{RQ}\textbf{Prompt 1: }You are an expert agile estimation assistant specialized in story point estimation for software development.

Item: \{$x$\}

OUTPUT FORMAT: \{"Story point": \}
\end{RQ}

\noindent For each backlog item $x_i$, its title and description are inputted as the \{title\} and \{description\} in the prompt and the LLM's output story point number will be recorded as $c^{(1)}_i\in C^{(1)}$.

The performance of the LLM is evaluated using the Pearson correlation coefficient $\rho$ and Spearman's rank correlation coefficient $r_s$ between its predictions $c_i\in C$ and the ground truth story points $y_i \in Y$:
\begin{equation}\label{rho}
    \rho(C,Y)=\frac{\sum\limits_{i=1}^{n}(c_i-{\overline{C})(y_{i}-{\overline {Y}})}}{\sqrt {\sum\limits_{i=1}^{n}(c_i-{\overline{C})^{2}}}{\sqrt {\sum\limits_{i=1}^{n}(y_{i}-{\overline {Y}})^{2}}}}
\end{equation}
\begin{equation}\label{rs}
    r_{s}(C,Y)={\rho } {\bigl (}\ {R} [C],{R} [Y]\ {\bigr )}
\end{equation}
where $R[C]$ and $R[Y]$ are the ranks of $C$ and $Y$. A higher $\rho(C,Y)$ suggests that the predicted story points $C$ have a better linear relationship with the ground truth story points $Y$. A higher $r_{s}(C,Y)$ indicates that the order of $C$ better matches that of $Y$, regardless of whether a linear relationship between $C$ and $Y$ exists.

By comparing the performance of LLMs in $\rho$ and $r_s$ with the performance of the baseline SBERT regression model~\cite{khan2025efficientstorypointestimation} trained on $80\%$ of the data, we can determine whether LLMs can achieve good performance without any training data.

\subsection{RQ2: How do LLMs perform in predicting the story points when provided with a few example story points?}

To explore this research question, we applied the following few-shot prompt to the four LLMs:

\begin{RQ}\textbf{Prompt 2: }You are an expert agile estimation assistant specialized in story point estimation for software development.

\{few-shot examples: 

Item: \{$x_i$\}  Story point: \{$y_i$\}\}

Based on these examples, please estimate the story points for the following issue:

Item: \{$x$\}

OUTPUT FORMAT:

\{"Story point": \}
\end{RQ}
\noindent For each project, five backlog items are selected from the training split as few-shot examples. These examples are provided to the LLM alongside their ground-truth story points. The LLM then predicts the story points of remaining backlog items $c^{(2)}_i\in C^{(2)}$. We investigate two strategies for selecting these five examples:

\bi
\item \textbf{Prompt 2-Count} (\textit{Count-based selection}): Examples are selected by choosing one representative item from each of the five most frequently occurring story point values in the project's training split. This strategy reflects the empirical distribution of story points in the project, ensuring that the LLM is exposed to the most commonly assigned effort levels. For each of the top-five most frequent values, one item is sampled uniformly at random. Because this selection is driven by frequency, the examples may cluster in a narrow range of values if the project's story point distribution is skewed (e.g., dominated by low-effort items), giving the LLM limited information about the extremes of the scale.

\item \textbf{Prompt 2-Scale} (\textit{Scale-aware selection}): Examples are selected to maximize coverage of the project's story point scale. Specifically, the full range $[\min(Y), \max(Y)]$ is partitioned into five equal-width intervals, and one item is sampled from each interval. This ensures that the five demonstrated examples include at least one low-effort item, one high-effort item, and items spread across the middle bins. By explicitly spanning the estimation scale, this strategy provides the LLM with calibration anchors at multiple effort levels, helping it map its internal difficulty representations onto the project-specific story point range.
\ei

\noindent Both strategies fix the number of examples at five, so they are directly comparable in terms of annotation cost. The few-shot performance of both variants is compared with the zero-shot baseline (Prompt 1) in terms of $\rho(C,Y)$ and $r_s(C,Y)$ to see whether providing the project-specific few-shot examples helps the LLMs calibrate its story point predictions. The comparison between Prompt 2-Count and Prompt 2-Scale further reveals whether scale diversity or frequency representativeness is the more beneficial principle for few-shot example selection.

\subsection{RQ3: Is it easier for the LLMs to predict the comparative judgments than the story points?}

It has been shown, both theoretically~\cite{thurstone1927law} and empirically~\cite{khan2025efficientstorypointestimation}, that comparative judgments are easier to make for human developers than direct story point estimates. To test whether this also applies to LLMs, we compared the performance of the zero-shot prompt in RQ1 with the following comparative prediction prompt:

\begin{RQ}\textbf{Prompt 3: }You are an expert agile estimation assistant specialized in story point estimation for software development.

Item A: \{$x_i$\} 

Item B: \{$x_j$\}

Which issue has higher story points? Output 1 if A > B; -1 if B > A.

OUTPUT FORMAT:

\{"Decision": <1 or -1>\}
\end{RQ}

\noindent For each project of $n$ backlog items, $n$ pairs of items $(x_i, x_j)$ are randomly generated under the condition that $y_i\neq y_j$. The ground truth comparative judgments for these $n$ item pairs are $y_{ij} = \text{sgn}(y_i-y_j)\in Y_{ij}$, as shown in Figure~\ref{fig:data_gen} (b). The LLM's predicted decisions are recorded as $c^{(3)}_{ij}\in C^{(3)}_{ij}$. The performance of the LLM was evaluated using accuracy in \eqref{acc}.
\begin{equation}\label{acc}
    \text{Acc}(C_{ij},Y_{ij})=\frac{|C_{ij}=Y_{ij}|}{n}.
\end{equation}
Similarly, the accuracy of Prompt 1 from RQ1 is evaluated by calculating LLM's decisions as $c^{(1)}_{ij} = \text{sgn}(c^{(1)}_i-c^{(1)}_j) \in C^{(1)}_{ij}$. If it is easier for LLMs to predict comparative judgments than story points, we should be able to see $\text{Acc}(C^{(3)}_{ij},Y_{ij})>\text{Acc}(C^{(1)}_{ij},Y_{ij})$.

\subsection{RQ4: How do LLMs perform in predicting the story points when provided with a few comparative judgments as examples?}

In this research question, we tested the performance of the following few-shot prompt using comparative judgments as the few-shot examples:

\begin{RQ}\textbf{Prompt 4: }You are an expert agile estimation assistant specialized in story point estimation for software development.

\{few-shot examples: 

Item A: \{$x_i$\} Item B: \{$x_j$\} Decision: \{$y_{ij}$\}\}

Now, based on these examples, please estimate the story points for the following issue:

Item: \{$x$\}

OUTPUT FORMAT:

\{"Story point": \}
\end{RQ}

\begin{table*}[!tbh]
\caption{Results for RQ1. The zero-shot Prompt 1 results of the four LLMs are compared to the state-of-the-art deep learning results of Khan et al.~\cite{khan2025efficientstorypointestimation} trained on $80\%$ of the labeled data (Regression) or comparative judgment data (Comparative).}\label{tbl:RQ1}
\small
\centering
\setlength\tabcolsep{10pt}
\begin{tabular}{l|l|cc|cccc|}
\multirow{2}{*}{Metric}   & \multirow{2}{*}{Project}            & \multicolumn{2}{c|}{Khan et al.~\cite{khan2025efficientstorypointestimation}}  & \multicolumn{4}{c|}{Prompt 1}  \\\cline{3-8}
&& Regression & Comparative & Kimi & DeepSeek & Gemini & OpenAI \\\hline
\multirow{17}{*}{$\rho$}  
         & appceleratorstudio & 0.3254           & 0.333             & 0.3355        & 0.3392            & 0.3233          & 0.2718          \\
         & aptanastudio       & 0.3419           & 0.3452            & 0.3631        & 0.4222            & 0.1608          & 0.322           \\
         & bamboo             & 0.1768           & 0.186             & 0.316         & 0.4107            & 0.1608          & 0.2096          \\
         & clover             & 0.4403           & 0.419             & 0.5723        & 0.5208            & 0.1929          & 0.4096          \\
         & datamanagement     & 0.3775           & 0.3271            & 0.4596        & 0.4055            & 0.0212          & 0.15            \\
         & duracloud          & 0.3758           & 0.3858            & 0.5257        & 0.4659            & 0.4399          & 0.3968          \\
         & jirasoftware       & 0.5324           & 0.4414            & 0.2285        & 0.3203            & 0.1784          & 0.1882          \\
         & mesos              & 0.396            & 0.4359            & 0.4042        & 0.4238            & 0.3779          & 0.042           \\
         & moodle             & 0.3499           & 0.2929            & 0.3748        & 0.4135            & 0.322           & 0.0353          \\
         & mule               & 0.2342           & 0.3188            & 0.3922        & 0.4198            & 0.3186          & 0.319           \\
         & mulestudio         & 0.1096           & 0.2265            & 0.232         & 0.1741            & 0.167           & 0.1392          \\
         & springxd           & 0.3982           & 0.4066            & 0.4114        & 0.4138            & 0.3407          & 0.3113          \\
         & talenddataquality  & 0.2892           & 0.2983            & 0.309         & 0.3522            & 0.0168          & 0.3161          \\
         & talendesb          & 0.3453           & 0.4189            & 0.4614        & 0.4               & 0.4586          & 0.3948          \\
         & titanium           & 0.1861           & 0.2098            & 0.3474        & 0.3568            & 0.0887          & 0.2606          \\
         & usergrid           & 0.2016           & 0.2945            & 0.2425        & 0.2666            & 0.2131          & 0.2638          \\
  \rowcolor{gray!25}  \cellcolor{white}    &  average           & 0.3175 &	0.3337 &	0.3735 &	\textbf{0.3816} &	0.2363 &	0.2519         \\\hline
\multirow{17}{*}{$r_s$} 
	    & appceleratorstudio & 0.3037           & 0.3222            & 0.3778        & 0.3686            & 0.3381          & 0.3049          \\
         & aptanastudio       & 0.283            & 0.2682            & 0.3184        & 0.3713            & 0.2524          & 0.2621          \\
         & bamboo             & 0.1753           & 0.1761            & 0.348         & 0.3213            & 0.2662          & 0.2471          \\
         & clover             & 0.4166           & 0.4483            & 0.5043        & 0.4496            & 0.3831          & 0.4846          \\
         & datamanagement     & 0.3909           & 0.3794            & 0.492         & 0.4703            & 0.4556          & 0.4215          \\
         & duracloud          & 0.4221           & 0.4006            & 0.5703        & 0.5027            & 0.4909          & 0.4729          \\
         & jirasoftware       & 0.4987           & 0.4386            & 0.3944        & 0.4015            & 0.3279          & 0.3246          \\
         & mesos              & 0.3916           & 0.4402            & 0.474         & 0.4549            & 0.4253          & 0.4034          \\
         & moodle             & 0.3552           & 0.3276            & 0.4158        & 0.4438            & 0.3101          & 0.36            \\
         & mule               & 0.2459           & 0.3235            & 0.4566        & 0.4414            & 0.3837          & 0.4003          \\
         & mulestudio         & 0.0612           & 0.2148            & 0.3634        & 0.2744            & 0.2575          & 0.2668          \\
         & springxd           & 0.3911           & 0.4041            & 0.4443        & 0.4282            & 0.3846          & 0.412           \\
         & talenddataquality  & 0.2985           & 0.2973            & 0.3168        & 0.3245            & 0.2225          & 0.3495          \\
         & talendesb          & 0.355            & 0.4528            & 0.5177        & 0.4502            & 0.5509          & 0.4495          \\
         & titanium           & 0.2264           & 0.2439            & 0.3508        & 0.3259            & 0.3342          & 0.2981          \\
         & usergrid           & 0.1981           & 0.3075            & 0.2322        & 0.2437            & 0.1752          & 0.2072          \\
        \rowcolor{gray!25}  \cellcolor{white}    & average & 0.3133 &	0.3403 &	\textbf{0.4111} &	0.3920 & 	0.3474 &	0.3540  \\\hline     
\end{tabular}
\end{table*}

\noindent For each project, five pairs of items are selected as the few-shot examples. These examples are input into the LLM along with their associated decisions $y_{ij}$. The ``Decision'' is described as ``Item A has a higher story point than Item B'' when $y_{ij}=1$ and ``Item A has a lower story point than Item B'' when $y_{ij}=-1$. Then, the LLM predicts the story points of other backlog items $c^{(4)}_i\in C^{(4)}$. The few-shot performance is evaluated with $\rho(C^{(4)},Y)$ and $r_s(C^{(4)},Y)$. According to the results of Khan et al.~\cite{khan2025efficientstorypointestimation}, machine learning models can easily learn from comparative judgments. If the same conclusion holds for LLMs, we should see $\rho(C^{(4)},Y)>\rho(C^{(1)},Y)$ and $r_s(C^{(4)},Y)>r_s(C^{(1)},Y)$. Furthermore, we compared the performances of Prompts 2 and 4. If $\rho(C^{(4)},Y)\ge\rho(C^{(2)},Y)$ and $r_s(C^{(4)},Y)\ge r_s(C^{(2)},Y)$, providing comparative judgments as few-shot examples is beneficial to the LLMs since they are easier to acquire.

\section{Experimental Results}

\subsection{RQ1: How do LLMs perform in predicting the story points when there are no training data?}

Table~\ref{tbl:RQ1} presents the results of the four LLMs using the Prompt 1 and compares them with the state-of-the-art baseline results from Khan et al.~\cite{khan2025efficientstorypointestimation}. The baseline regression model was trained on $80\%$ of the story point data. The baseline comparative model was trained on pairs generated within the $80\%$ training data and thus require human annotations for the comparative judgments. Compared to these two baseline models, Prompt 1 used no training data at all and thus require zero human effort in annotation. From Table~\ref{tbl:RQ1}, we can observe:
\bi
\item
Kimi and DeepSeek performed the best. While DeepSeek got the highest $\rho=0.3816$ on average, Kimi achieved the highest $r_s=0.4111$ on average. 
\item
Although no training data is used, Kimi and DeepSeek outperformed the best supervised deep learning baseline models in terms of both $\rho$ and $r_s$.
\item
Gemini and OpenAI achieved lower $\rho$ but higher $r_s$ when compared to the baseline models. This is likely because these two LLMs struggled with the correct scales of story points for different projects but still correctly predict the orders/ranks.
\ei

\noindent\textbf{Answer to RQ1: }Overall, using zero-shot prompts, LLMs are capable of predicting story points, especially the relative ranks of the story points, better than the supervised deep learning models. Kimi and DeepSeek performed better than Gemini and OpenAI in this task.

\begin{table*}[!tbh]
\caption{Results for RQ2. The zero-shot Prompt 1 results are compared against two versions of the few-shot Prompt 2 results of the four LLMs.}\label{tbl:RQ2}
\small
\centering
\setlength\tabcolsep{2pt}
\begin{tabular}{l|l|cp{1cm}p{1.1cm}|cp{1cm}p{1.1cm}|cp{1cm}p{1.1cm}|cp{1cm}p{1.1cm}|}
\multirow{2}{*}{Metric}   & \multirow{2}{*}{Project}            & \multicolumn{3}{c|}{Kimi}  & \multicolumn{3}{c|}{DeepSeek} & \multicolumn{3}{c|}{Gemini} & \multicolumn{3}{c|}{OpenAI} \\\cline{3-14}
& & Prompt 1 & Prompt 2-Scale & Prompt 2-Count & Prompt 1 & Prompt 2-Scale & Prompt 2-Count & Prompt 1 & Prompt 2-Scale & Prompt 2-Count & Prompt 1 & Prompt 2-Scale & Prompt 2-Count \\\hline
\multirow{17}{*}{$\rho$}  
         & appceleratorstudio & 0.3355        & 0.3481              & 0.3052              & 0.3392            & 0.3394                  & 0.2877                  & 0.3233          & 0.3227                & 0.3297                & 0.2718          & 0.3020                & 0.3063                \\
         & aptanastudio       & 0.3631        & 0.4557              & 0.3895              & 0.4222            & 0.4652                  & 0.3679                  & 0.1608          & 0.2974                & 0.3396                & 0.3220          & 0.3725                & 0.3408                \\
         & bamboo             & 0.3160        & 0.5236              & 0.3326              & 0.4107            & 0.4704                  & 0.3869                  & 0.1608          & 0.1139                & 0.2605                & 0.2096          & 0.1711                & 0.2071                \\
         & clover             & 0.5723        & 0.5796              & 0.5935              & 0.5208            & 0.6095                  & 0.5344                  & 0.1929          & 0.3660                & 0.4629                & 0.4096          & 0.4248                & 0.4638                \\
         & datamanagement     & 0.4596        & 0.5468              & 0.4494              & 0.4055            & 0.4583                  & 0.4658                  & 0.0212          & 0.4152                & 0.2783                & 0.1500          & 0.3692                & 0.3158                \\
         & duracloud          & 0.5257        & 0.5337              & 0.5263              & 0.4659            & 0.5866                  & 0.5507                  & 0.4399          & 0.4409                & 0.4206                & 0.3968          & 0.4300                & 0.3891                \\
         & jirasoftware       & 0.2285        & 0.4280              & 0.2998              & 0.3203            & 0.4436                  & 0.3242                  & 0.1784          & 0.3287                & 0.1983                & 0.1882          & 0.1844                & 0.2165                \\
         & mesos              & 0.4042        & 0.5013              & 0.4810              & 0.4238            & 0.4908                  & 0.4413                  & 0.3779          & 0.3815                & 0.3196                & 0.0420          & 0.4021                & 0.4128                \\
         & moodle             & 0.3748        & 0.4375              & 0.4419              & 0.4135            & 0.4502                  & 0.4750                  & 0.3220          & 0.3438                & 0.3439                & 0.0353          & 0.4214                & 0.3902                \\
         & mule               & 0.3922        & 0.4500              & 0.4726              & 0.4198            & 0.5038                  & 0.4795                  & 0.3186          & 0.3129                & 0.3594                & 0.3190          & 0.3483                & 0.4082                \\
         & mulestudio         & 0.2320        & 0.1606              & 0.1888              & 0.1741            & 0.2503                  & 0.1964                  & 0.1670          & 0.1370                & 0.1949                & 0.1392          & 0.1564                & 0.1912                \\
         & springxd           & 0.4114        & 0.4603              & 0.4306              & 0.4138            & 0.4771                  & 0.4179                  & 0.3407          & 0.3399                & 0.3355                & 0.3113          & 0.3501                & 0.3649                \\
         & talenddataquality  & 0.3090        & 0.3577              & 0.3575              & 0.3522            & 0.3822                  & 0.3391                  & 0.0168          & 0.2389                & 0.3130                & 0.3161          & 0.3481                & 0.3538                \\
         & talendesb          & 0.4614        & 0.4415              & 0.4457              & 0.4000            & 0.4968                  & 0.4730                  & 0.4586          & 0.3686                & 0.4105                & 0.3948          & 0.3822                & 0.4035                \\
         & titanium           & 0.3474        & 0.4207              & 0.4093              & 0.3568            & 0.4078                  & 0.4186                  & 0.0887          & 0.2640                & 0.2366                & 0.2606          & 0.1636                & 0.1624                \\
         & usergrid           & 0.2425        & 0.3261              & 0.3625              & 0.2666            & 0.4194                  & 0.3845                  & 0.2131          & 0.3822                & 0.3230                & 0.2638          & 0.2731                & 0.2260                \\
 \rowcolor{gray!25} \cellcolor{white}       & average            & 0.3735     & \textbf{0.4357}              & 0.4054              & 0.3816            & \textbf{0.4532}                  & 0.4089                  & 0.2363          & 0.3158                & \textbf{0.3204}                & 0.2519          & 0.3187                & \textbf{0.3220}                \\ 
 \hline
\multirow{17}{*}{$r_s$} 
         & appceleratorstudio & 0.3778        & 0.4063              & 0.3422              & 0.3686            & 0.3883                  & 0.3274                  & 0.3381          & 0.3274                & 0.3592                & 0.3049          & 0.3790                & 0.3415                \\
         & aptanastudio       & 0.3184        & 0.4117              & 0.3846              & 0.3713            & 0.3683                  & 0.3618                  & 0.2524          & 0.3618                & 0.3507                & 0.2621          & 0.2994                & 0.2707                \\
         & bamboo             & 0.3480        & 0.3877              & 0.4054              & 0.3213            & 0.3529                  & 0.3338                  & 0.2662          & 0.3338                & 0.3154                & 0.2471          & 0.2535                & 0.3154                \\
         & clover             & 0.5043        & 0.5480              & 0.5596              & 0.4496            & 0.5395                  & 0.5531                  & 0.3831          & 0.5531                & 0.4837                & 0.4846          & 0.4604                & 0.4835                \\
         & datamanagement     & 0.4920        & 0.5739              & 0.4837              & 0.4703            & 0.5447                  & 0.4990                  & 0.4556          & 0.4990                & 0.3335                & 0.4215          & 0.3913                & 0.3659                \\
         & duracloud          & 0.5703        & 0.5948              & 0.6031              & 0.5027            & 0.5856                  & 0.5589                  & 0.4909          & 0.5589                & 0.4766                & 0.4729          & 0.5330                & 0.4952                \\
         & jirasoftware       & 0.3944        & 0.4970              & 0.4318              & 0.4015            & 0.4839                  & 0.4045                  & 0.3279          & 0.4045                & 0.3108                & 0.3246          & 0.3138                & 0.3775                \\
         & mesos              & 0.4740        & 0.5311              & 0.5408              & 0.4549            & 0.5145                  & 0.5123                  & 0.4253          & 0.5123                & 0.3404                & 0.4034          & 0.4488                & 0.4422                \\
         & moodle             & 0.4158        & 0.4330              & 0.4163              & 0.4438            & 0.4690                  & 0.4713                  & 0.3101          & 0.4713                & 0.3048                & 0.3600          & 0.3347                & 0.3336                \\
         & mule               & 0.4566        & 0.4515              & 0.4544              & 0.4414            & 0.4709                  & 0.4901                  & 0.3837          & 0.4901                & 0.3435                & 0.4003          & 0.3911                & 0.4189                \\
         & mulestudio         & 0.3634        & 0.2915              & 0.3137              & 0.2744            & 0.3426                  & 0.2996                  & 0.2575          & 0.2996                & 0.2998                & 0.2668          & 0.3311                & 0.3222                \\
         & springxd           & 0.4443        & 0.4957              & 0.4895              & 0.4282            & 0.4921                  & 0.4817                  & 0.3846          & 0.4817                & 0.3922                & 0.4120          & 0.4438                & 0.4465                \\
         & talenddataquality  & 0.3168        & 0.3825              & 0.4143              & 0.3245            & 0.3658                  & 0.3630                  & 0.2225          & 0.3630                & 0.3328                & 0.3495          & 0.3021                & 0.3701                \\
         & talendesb          & 0.5177        & 0.4965              & 0.5153              & 0.4502            & 0.5126                  & 0.5095                  & 0.5509          & 0.5095                & 0.4804                & 0.4495          & 0.4605                & 0.4558                \\
         & titanium           & 0.3508        & 0.4178              & 0.3948              & 0.3259            & 0.3984                  & 0.3889                  & 0.3342          & 0.3889                & 0.2365                & 0.2981          & 0.3744                & 0.3232                \\
         & usergrid           & 0.2322        & 0.3167              & 0.3256              & 0.2437            & 0.3828                  & 0.3073                  & 0.1752          & 0.3073                & 0.2849                & 0.2072          & 0.2759                & 0.1896                \\
    \rowcolor{gray!25} \cellcolor{white}       & average            & 0.4111       & \textbf{0.4522}         & 0.4422          & 0.3920          &\textbf{0.4507}              & 0.4289                & 0.3474       & \textbf{0.4289}            & 0.3528              & 0.3540      & \textbf{0.3745 }           & 0.3720            \\\hline
\end{tabular}
\end{table*}

\subsection{RQ2: How do LLMs perform in predicting the story points when provided with a few example story points?}

Table~\ref{tbl:RQ2} presents the few-shot performance of the four LLMs under Prompt 2 and compares both selection strategies against the zero-shot Prompt 1 baseline. From Table~\ref{tbl:RQ2}, we can observe:
\bi
\item
Few-shot prompting consistently improves average performance over the zero-shot baseline across all four LLMs. Although there are individual projects where the zero-shot baseline is not surpassed, the average improvement is clear. For DeepSeek and Kimi, the average $\rho$ increases from 0.3816 and 0.3735 (Prompt 1) to 0.4532 and 0.4357 (Prompt 2-Scale), respectively. Gemini and OpenAI exhibit similarly consistent average gains, improving from 0.2363 and 0.2519 to 0.3204 and 0.3220 in $\rho$. Even larger improvements have been observed for $r_s$.

\item
Prompt 2-Scale outperforms Prompt 2-Count on average across all four LLMs especially in terms of $r_s$. While this advantage is consistent at the average level, per-project variation exists, particularly for Spearman rank correlation where Prompt 2-Count occasionally performs better (e.g., in 9 of 16 projects for Kimi). However, overall, this indicates that covering a broader range of story point values in the few-shot examples is generally more beneficial than matching the most frequent values in the training distribution.

\item
Few-shot improvements are more pronounced for Gemini and OpenAI, which had lower zero-shot baselines. This suggests that providing explicit scale anchors through few-shot examples particularly benefits models that struggle with calibrating absolute story point values in the zero-shot setting.
\ei

\noindent\textbf{Answer to RQ2: }Overall, providing a small number of labeled examples substantially improves the average story point prediction performance of all four LLMs. The scale-aware selection strategy (Prompt 2-Scale) delivers stronger average gains than the count-based strategy, highlighting the importance of covering the full range of project-specific story point values in the few-shot examples.

\begin{table*}[!tbh]
\caption{Results for RQ3. This table presents the accuracy of Prompt 3 (direct comparative prediction) for the four LLMs. The pairwise accuracy from Prompt 1 is computed from its zero-shot story point predictions. Random baseline accuracy is 0.50.}\label{tbl:RQ3}
\small
\centering
\setlength\tabcolsep{5pt}
\begin{tabular}{l|l|p{1.42cm}p{1.42cm}|p{1.42cm}p{1.42cm}|p{1.42cm}p{1.42cm}|p{1.42cm}p{1.42cm}|}
\multirow{2}{*}{Metric} & \multirow{2}{*}{Project} & \multicolumn{2}{c|}{Kimi} & \multicolumn{2}{c|}{DeepSeek} & \multicolumn{2}{c|}{Gemini} & \multicolumn{2}{c|}{OpenAI} \\\cline{3-10}
 & & Prompt 1 & Prompt 3 & Prompt 1 & Prompt 3 & Prompt 1 & Prompt 3 & Prompt 1 & Prompt 3 \\\hline
\multirow{17}{*}{Acc.} & appceleratorstudio  & 0.6075 & 0.5697 & 0.7547 & 0.6207 & 0.5642 & 0.5478 & 0.6653 & 0.6336 \\
                       & aptanastudio        & 0.5750 & 0.5645 & 0.6576 & 0.5564 & 0.5875 & 0.5597 & 0.6429 & 0.5978 \\
                       & bamboo              & 0.6790 & 0.6123 & 0.7500 & 0.6176 & 0.5509 & 0.5125 & 0.5873 & 0.5582 \\
                       & clover              & 0.7199 & 0.6016 & 0.8377 & 0.7341 & 0.6094 & 0.5807 & 0.6510 & 0.6402 \\
                       & datamanagement      & 0.6925 & 0.5835 & 0.7643 & 0.5946 & 0.5858 & 0.5494 & 0.6222 & 0.5700 \\
                       & duracloud           & 0.6273 & 0.5075 & 0.8515 & 0.7341 & 0.6574 & 0.6557 & 0.7299 & 0.6727 \\
                       & jirasoftware        & 0.6923 & 0.6655 & 0.7557 & 0.6466 & 0.6373 & 0.5528 & 0.7113 & 0.5487 \\
                       & mesos               & 0.6723 & 0.5542 & 0.7766 & 0.6549 & 0.5804 & 0.5685 & 0.6792 & 0.6639 \\
                       & moodle              & 0.5904 & 0.5840 & 0.7357 & 0.6244 & 0.6312 & 0.5832 & 0.6690 & 0.6039 \\
                       & mule                & 0.6908 & 0.6052 & 0.7750 & 0.6232 & 0.5399 & 0.5163 & 0.6524 & 0.6074 \\
                       & mulestudio          & 0.6377 & 0.5956 & 0.6494 & 0.5902 & 0.6161 & 0.5615 & 0.6598 & 0.5916 \\
                       & springxd            & 0.6662 & 0.5964 & 0.7646 & 0.6625 & 0.5729 & 0.5559 & 0.6860 & 0.6536 \\
                       & talenddataquality   & 0.6893 & 0.5851 & 0.6853 & 0.5414 & 0.5592 & 0.5223 & 0.6104 & 0.5574 \\
                       & talendesb           & 0.5790 & 0.5242 & 0.7830 & 0.6968 & 0.5484 & 0.5357 & 0.6982 & 0.6614 \\
                       & titanium            & 0.6685 & 0.6024 & 0.7133 & 0.5891 & 0.5096 & 0.5002 & 0.5731 & 0.5403 \\
                       & usergrid            & 0.5895 & 0.5529 & 0.6284 & 0.6036 & 0.6369 & 0.5967 & 0.6900 & 0.5913 \\
\rowcolor{gray!25} \cellcolor{white} & average & \textbf{0.6486} & 0.5815 & \textbf{0.7427} & 0.6306 & \textbf{0.5867} & 0.5562 & \textbf{0.6580} & 0.6058 \\\hline
\end{tabular}
\end{table*}

\subsection{RQ3: Is it easier for the LLMs to predict the comparative judgments than the story points?}

Table~\ref{tbl:RQ3} presents the per-project accuracy of all four LLMs under Prompt 3 (direct zero-shot comparative prediction). For reference, the derived accuracy from Prompt 1 is also shown, computed as $\text{Acc}(C^{(1)}_{ij}, Y_{ij})$ using the story point predictions from the zero-shot setting $c^{(1)}_{ij} = \text{sgn}(c^{(1)}_i-c^{(1)}_j) \in C^{(1)}_{ij}$. From Table~\ref{tbl:RQ3}, we can observe:
\bi
\item
All four LLMs achieve average pairwise accuracy above the random baseline of 0.50 under Prompt 3. DeepSeek achieves the highest average accuracy of 0.6306, while OpenAI averages 0.6058, Kimi averages 0.5815, and Gemini averages 0.5562. 
\item
However, the derived accuracy from Prompt 1 exceeds the accuracy from Prompt 3 for all models. DeepSeek's derived accuracy (0.7427) substantially surpasses its Prompt 3 accuracy (0.6306), and Kimi (0.6486 vs. 0.5815), Gemini (0.5867 vs. 0.5562), and OpenAI (0.6580 vs. 0.6058) show the same pattern. This indicates that LLMs produce implicit pairwise orderings more reliably when predicting story points directly than when asked to compare items explicitly.
\item
Unlike in human subject studies where pairwise comparisons are considered more natural and less cognitively demanding than absolute estimation~\cite{thurstone1927law}, LLMs do not find it easier to compare two items than evaluating one item at a time. The gap between derived and direct comparative accuracy suggests that LLMs may internally rely on a latent numerical representation even when generating comparative decisions.
\ei

\noindent\textbf{Answer to RQ3: }It is not easier for LLMs to provide comparative judgments between a pair of items than predicting the story points for each one of the item. Direct zero-shot story point predictions yield more accurate implied pairwise orderings than explicit comparative predictions across all four LLMs, suggesting that LLMs approach effort estimation through a fundamentally different process than human annotators.

\begin{table*}[!tbh]
\caption{Results for RQ4. The comparative few-shot Prompt 4 results are compared against the Prompt 1 and Prompt 2 results of the four LLMs.}\label{tbl:RQ4}
\small
\centering
\setlength\tabcolsep{3pt}
\begin{tabular}{l|l|p{1cm}p{1cm}p{1cm}|p{1cm}p{1cm}p{1cm}|p{1cm}p{1cm}p{1cm}|p{1cm}p{1cm}p{1cm}|}
\multirow{2}{*}{Metric}  & \multirow{2}{*}{Project} & \multicolumn{3}{c|}{Kimi}                                  & \multicolumn{3}{c|}{DeepSeek}                              & \multicolumn{3}{c|}{Gemini}                                & \multicolumn{3}{c|}{OpenAI}                                \\ \cline{3-14} 
                         &                          & Prompt 1 & Prompt 2        & Prompt 4 & Prompt 1 & Prompt 2        & Prompt 4 & Prompt 1 & Prompt 2        & Prompt 4 & Prompt 1 & Prompt 2        & Prompt 4 \\ \hline
\multirow{17}{*}{$\rho$}  
	 & appceleratorstudio & 0.3355        & 0.3481        & 0.3035                  & 0.3392        & 0.3394        & 0.3037                  & 0.3233        & 0.3227        & 0.3510                  & 0.2718          & 0.3020        & 0.3020          \\
         & aptanastudio       & 0.3631        & 0.4557        & 0.3949                  & 0.4222        & 0.4652        & 0.4356                  & 0.1608        & 0.2974        & 0.3624                  & 0.3220          & 0.3725        & 0.3384          \\
         & bamboo             & 0.3160        & 0.5236        & 0.3745                  & 0.4107        & 0.4704        & 0.5357                  & 0.1608        & 0.1139        & 0.2519                  & 0.2096          & 0.1711        & 0.1514          \\
         & clover             & 0.5723        & 0.5796        & 0.3923                  & 0.5208        & 0.6095        & 0.5882                  & 0.1929        & 0.3660        & 0.5173                  & 0.4096          & 0.4248        & 0.3762          \\
         & datamanagement     & 0.4596        & 0.5468        & 0.4560                  & 0.4055        & 0.4583        & 0.4811                  & 0.0212        & 0.4152        & 0.2902                  & 0.1500          & 0.3692        & 0.3463          \\
         & duracloud          & 0.5257        & 0.5337        & 0.6037                  & 0.4659        & 0.5866        & 0.5853                  & 0.4399        & 0.4409        & 0.5687                  & 0.3968          & 0.4300        & 0.4829          \\
         & jirasoftware       & 0.2285        & 0.4280        & 0.3073                  & 0.3203        & 0.4436        & 0.2921                  & 0.1784        & 0.3287        & 0.2447                  & 0.1882          & 0.1844        & 0.2490          \\
         & mesos              & 0.4042        & 0.5013        & 0.4548                  & 0.4238        & 0.4908        & 0.4661                  & 0.3779        & 0.3815        & 0.4625                  & 0.0420          & 0.4021        & 0.3760          \\
         & moodle             & 0.3748        & 0.4375        & 0.4285                  & 0.4135        & 0.4502        & 0.4925                  & 0.3220        & 0.3438        & 0.4358                  & 0.0353          & 0.4214        & 0.3829          \\
         & mule               & 0.3922        & 0.4500        & 0.4549                  & 0.4198        & 0.5038        & 0.4458                  & 0.3186        & 0.3129        & 0.3911                  & 0.3190          & 0.3483        & 0.3204          \\
         & mulestudio         & 0.2320        & 0.1606        & 0.2111                  & 0.1741        & 0.2503        & 0.2100                  & 0.1670        & 0.1370        & 0.1932                  & 0.1392          & 0.1564        & 0.1844          \\
         & springxd           & 0.4114        & 0.4603        & 0.3770                  & 0.4138        & 0.4771        & 0.4849                  & 0.3407        & 0.3399        & 0.4513                  & 0.3113          & 0.3501        & 0.3537          \\
         & talenddataquality  & 0.3090        & 0.3577        & 0.3526                  & 0.3522        & 0.3822        & 0.3198                  & 0.0168        & 0.2389        & 0.3934                  & 0.3161          & 0.3481        & 0.3101          \\
         & talendesb          & 0.4614        & 0.4415        & 0.4698                  & 0.4000        & 0.4968        & 0.5420                  & 0.4586        & 0.3686        & 0.4427                  & 0.3948          & 0.3822        & 0.4059          \\
         & titanium           & 0.3474        & 0.4207        & 0.3434                  & 0.3568        & 0.4078        & 0.4156                  & 0.0887        & 0.2640        & 0.3366                  & 0.2606          & 0.1636        & 0.2900          \\
         & usergrid           & 0.2425        & 0.3261        & 0.3020                  & 0.2666        & 0.4194        & 0.3015                  & 0.2131        & 0.3822        & 0.3270                  & 0.2638          & 0.2731        & 0.2908          \\
      \rowcolor{gray!25} \cellcolor{white}                   
	 & average            & 0.3735   & \textbf{0.4357}    & 0.3891                  & 0.3816    & \textbf{0.4532}    & 0.4312                  & 0.2363        & 0.3158   & \textbf{0.3762}              & 0.2519        & 0.3187     & \textbf{0.3225}       \\ 
\hline
\multirow{17}{*}{$r_s$}  
	 & appceleratorstudio & 0.3778        & 0.4063        & 0.3412                  & 0.3686        & 0.3883        & 0.3468                  & 0.3381        & 0.3274        & 0.3593                  & 0.3049          & 0.3790          & 0.3050          \\
         & aptanastudio       & 0.3184        & 0.4117        & 0.3943                  & 0.3713        & 0.3683        & 0.3469                  & 0.2524        & 0.3618        & 0.3188                  & 0.2621          & 0.2994          & 0.2802          \\
         & bamboo             & 0.3480        & 0.3877        & 0.3910                  & 0.3213        & 0.3529        & 0.4349                  & 0.2662        & 0.3338        & 0.3160                  & 0.2471          & 0.2535          & 0.1864          \\
         & clover             & 0.5043        & 0.5480        & 0.5163                  & 0.4496        & 0.5395        & 0.5759                  & 0.3831        & 0.5531        & 0.5059                  & 0.4846          & 0.4604          & 0.4041          \\
         & datamanagement     & 0.4920        & 0.5739        & 0.5230                  & 0.4703        & 0.5447        & 0.5040                  & 0.4556        & 0.4990        & 0.3717                  & 0.4215          & 0.3913          & 0.4374          \\
         & duracloud          & 0.5703        & 0.5948        & 0.5875                  & 0.5027        & 0.5856        & 0.5855                  & 0.4909        & 0.5589        & 0.5822                  & 0.4729          & 0.5330          & 0.5090          \\
         & jirasoftware       & 0.3944        & 0.4970        & 0.4581                  & 0.4015        & 0.4839        & 0.4216                  & 0.3279        & 0.4045        & 0.3450                  & 0.3246          & 0.3138          & 0.3860          \\
         & mesos              & 0.4740        & 0.5311        & 0.5424                  & 0.4549        & 0.5145        & 0.5092                  & 0.4253        & 0.5123        & 0.4883                  & 0.4034          & 0.4488          & 0.4374          \\
         & moodle             & 0.4158        & 0.4330        & 0.3913                  & 0.4438        & 0.4690        & 0.4088                  & 0.3101        & 0.4713        & 0.3976                  & 0.3600          & 0.3347          & 0.3675          \\
         & mule               & 0.4566        & 0.4515        & 0.4702                  & 0.4414        & 0.4709        & 0.4471                  & 0.3837        & 0.4901        & 0.3985                  & 0.4003          & 0.3911          & 0.3514          \\
         & mulestudio         & 0.3634        & 0.2915        & 0.3283                  & 0.2744        & 0.3426        & 0.3236                  & 0.2575        & 0.2996        & 0.3053                  & 0.2668          & 0.3311          & 0.2756          \\
         & springxd           & 0.4443        & 0.4957        & 0.4697                  & 0.4282        & 0.4921        & 0.4932                  & 0.3846        & 0.4817        & 0.4726                  & 0.4120          & 0.4438          & 0.3960          \\
         & talenddataquality  & 0.3168        & 0.3825        & 0.4054                  & 0.3245        & 0.3658        & 0.3131                  & 0.2225        & 0.3630        & 0.3935                  & 0.3495          & 0.3021          & 0.3027          \\
         & talendesb          & 0.5177        & 0.4965        & 0.5216                  & 0.4502        & 0.5126        & 0.5784                  & 0.5509        & 0.5095        & 0.4925                  & 0.4495          & 0.4605          & 0.4750          \\
         & titanium           & 0.3508        & 0.4178        & 0.3685                  & 0.3259        & 0.3984        & 0.3490                  & 0.3342        & 0.3889        & 0.3271                  & 0.2981          & 0.3744          & 0.2635          \\
         & usergrid           & 0.2322        & 0.3167        & 0.3319                  & 0.2437        & 0.3828        & 0.2997                  & 0.1752        & 0.3073        & 0.3120                  & 0.2072          & 0.2759          & 0.2931          \\
           \rowcolor{gray!25} \cellcolor{white}                
	 & average            & 0.4111   & \textbf{0.4522}    & 0.4400                  & 0.3920   & \textbf{0.4507}    & 0.4336                  & 0.3474    & \textbf{0.4289}   & 0.3991                  & 0.3540     & \textbf{0.3745}      & 0.3544           \\ \hline
\end{tabular}
\end{table*}

\subsection{RQ4: How do LLMs perform in predicting the story points when provided with a few comparative judgments as examples?}

Table~\ref{tbl:RQ4} presents the performance of all four LLMs under Prompt 4 and compares it with Prompt 1 (zero-shot) and Prompt 2 (the best-performing scale-aware few-shot variant). From Table~\ref{tbl:RQ4}, we can observe:
\bi
\item
Comparative few-shot prompting (Prompt 4) improves average prediction performance over the zero-shot baseline across all four LLMs. In terms of $r_s$, all four models show positive average gains, ranging from negligible ($+0.0004$ for OpenAI) to substantial ($+0.0517$ for Gemini); however, per-project results are mixed for models such as DeepSeek, where Prompt 4 underperforms Prompt 1 in 8 of 16 projects. For $\rho$, all four models improve on average---gains of 0.016 (Kimi), 0.050 (DeepSeek), 0.051 (OpenAI), and 0.140 (Gemini). Overall, comparative examples carry sufficient ordinal information to calibrate the LLMs' absolute predictions.
\item
For DeepSeek and Kimi, Prompt 2-Scale still outperforms Prompt 4, indicating that direct numerical examples remain the stronger form of few-shot supervision for capable models. The difference in average $\rho$ is 0.022 for DeepSeek and 0.0466 for Kimi, suggesting that explicit story point labels provide more precise scale calibration than relative comparisons.
\item
For OpenAI and Gemini, Prompt 4 achieves higher average $\rho$ but lower average $r_s$ than Prompt 2-Scale. This suggests that for models with weaker absolute scale calibration, the simplified relative signal provided by comparative judgments can be more informative than a small set of labeled story points.
\ei

\noindent\textbf{Answer to RQ4: }Comparative judgments can serve as effective few-shot supervision signals for LLM-based story point estimation. Although direct labeled examples yield higher performance on average for Kimi, DeepSeek, and OpenAI, Prompt 4 produces positive average gains over the zero-shot baseline for all four models in both $\rho$ and $r_s$. Notably, Prompt 4 outperforms Prompt 2-Scale for Gemini and OpenAI $\rho$, demonstrating that pairwise comparative examples---which require substantially less annotation effort per example than labeled story points---are a practical and effective alternative source of few-shot supervision.

\section{Discussions}
\subsection{Effectiveness of Zero-shot and Few-shot Prompting}
Across all evaluated models (DeepSeek, Kimi, Gemini, and OpenAI), zero-shot prompting achieves moderate correlations, demonstrating that large language models possess transferable prior knowledge relevant to effort estimation. Even without project-specific training data, models are able to extract complexity signals from issue titles and descriptions.

However, few-shot prompting consistently improves performance across all models. For DeepSeek, the average Pearson correlation increases from approximately 0.3816 in zero-shot to 0.4532 under standard few-shot prompting. Kimi, Gemini, and OpenAI show similar positive trends, with few-shot settings outperforming their respective zero-shot baselines.

This pattern suggests that a small number of labeled examples is beneficial to calibrate models toward project-specific estimation scales. In particular, strategies that cover a broader range of story point scales (e.g., ScaleAware) tend to yield stronger improvements than frequency-based selection for capable models like DeepSeek, Kimi, and OpenAI, indicating that scale diversity is more beneficial than representativeness.

\subsection{Comparative Judgments vs. Direct Estimation For LLMs}
Contrary to findings in human subject studies, comparative judgment prediction is not inherently easier for LLMs than direct story point estimation.
In RQ3, zero-shot pairwise prompting achieves accuracies above random guessing for all models (DeepSeek: 0.631, OpenAI: 0.606, Kimi: 0.582, Gemini: 0.556), but these do not surpass the derived pairwise accuracy from direct zero-shot predictions (DeepSeek: 0.743, OpenAI: 0.658, Kimi: 0.649, Gemini: 0.587). This suggests that models may internally rely on a latent numerical representation even when producing comparative outputs.

\subsection{Comparative Judgments as Supervision Signals}
Although comparative judgments are not always easier to predict directly, they serve as effective supervision signals across the board.
Comparative few-shot prompting improves both Pearson and Spearman correlations over the zero-shot baseline for all four models. For Gemini specifically, it proved to be the most effective few-shot strategy, surpassing standard labeled examples. This demonstrates that relative supervision provides meaningful calibration information, even without exposing explicit numeric labels.



\subsection{Ranking vs. Absolute Calibration}
Across all experimental settings and models, Spearman correlations are consistently comparable to or higher than Pearson correlations. Kimi, for instance, achieves a Spearman correlation of 0.411 in zero-shot compared to a Pearson of 0.373. 
This indicates that LLMs are better at preserving relative ordering than matching exact numerical magnitudes. This observation aligns with the nature of story points: they are relative and project-specific rather than globally standardized numeric measures. LLMs appear to capture the relative difficulty of tasks more reliably than the exact scale used by a specific development team.

\subsection{Practical Implications}
From a practical perspective, these findings suggest that LLM-based story point estimation is viable in data-scarce environments, but the optimal strategy depends on the model used. 
Zero-shot performance provides a reasonable baseline. When limited supervision is available (e.g., five examples), direct few-shot with diverse examples (Scale/Count) is recommended for high-capacity models like DeepSeek, Kimi, and OpenAI. 
However, for smaller or more resource-constrained models like Gemini Flash Lite, comparative supervision offers a superior alternative. This opens the possibility of hybrid agile estimation workflows where teams provide relative comparisons—comparatively easier to collect—to effectively calibrate lighter-weight models.

\section{Threats to Validity}

\noindent \textbf{Conclusion validity: }Due to budget and time limits, there are not enough results for statistical tests. This poses a threat to the conclusion validity. However, because the differences in performance are not marginal, we still believe our conclusions are valid. Statistical tests will be applied with more repeated experiments in the future.

\noindent \textbf{Construct validity: }Because story points are unitless measurement of efforts, Pearson correlation coefficient $\rho$ is a good metric reflecting how well the LLM predicted story points linearly align with the human-estimated story points. In the meantime, Spearman's rank correlation coefficient $r_s$ reflects how well the order or rank of the LLM predicted story points align with the human-estimated story points.

\noindent \textbf{Internal validity: }Different design of the prompts might affect the LLMs' output. This poses a threat to the internal validity of this study. To reduce such threat, we have applied as simple prompt instructions as possible through Prompt 1 to 4.

\noindent \textbf{External validity: }This work is subject to threats to external validity. (1)~The choice of four LLMs may not be representative of the full population of LLMs. (2)~The experiments were conducted on 16 software projects; conclusions may not generalize to other domains or project types.

\section{Conclusion and Future Work}

In this study, we conducted a comprehensive evaluation of Large Language Models (LLMs) for agile story point estimation across 16 software projects. By benchmarking four distinct models—DeepSeek-V3.2, Kimi K2, Gemini Flash Lite, and GPT-5 Nano, we examined the effectiveness of zero-shot, few-shot, and comparative prompting strategies.

Our findings lead to four key conclusions:
\begin{itemize}
    \item \textbf{LLMs are effective estimators:} Even without fine-tuning, modern LLMs possess transferable knowledge that allows them to estimate effort with moderate correlation. Providing just five project-specific examples significantly boosts prediction and ranking performance, improving calibration toward project-specific story point scales.
    \item \textbf{Strategy matters:} Choosing \textit{diverse} few-shot examples (covering the full range of story points) is more effective than selecting the most frequent ones. This "ScaleAware" strategy proved robust across different models.
    \item \textbf{LLMs are different from humans in decision making:} While comparative judgments are easier for humans to provide than direct estimates~\cite{thurstone1927law}, they are not easier for the LLMs to predict. This suggests that the decision-making process of LLMs are different from that of human beings in nature.
    \item \textbf{One size does not fit all:} While high-capacity models like DeepSeek and Kimi perform best with direct labeled examples, resource-constrained models like Gemini Flash Lite benefit significantly more from comparative supervision. This suggests that relative judgments can serve as a powerful "scaffold" for smaller models.
\end{itemize}

We plan to broaden the scope of this study in future work:
\be
\item
Currently, our models rely on titles and descriptions; incorporating developer comments, acceptance criteria, and historical pull requests could further enhance accuracy. 
\item
Additionally, we aim to explore \textit{human-in-the-loop estimation workflows}, where human experts provide a small set of pairwise comparisons to calibrate a lightweight local model, thereby combining human intuition with automated scalability. 
\item
Next, investigating the potential of Chain-of-Thought (CoT) prompting to explicate the reasoning behind complex estimation tasks remains a promising avenue.
\item
Finally, we aim to explore whether more training data would further improve the LLMs' prediction performance. Supervised fine-tuning and reinforcement fine-tuning will be applied to train the LLMs.
\ee

\section*{Acknowledgement}
This work was funded by NSF Grants 2245796. 


\bibliographystyle{IEEEtran}
\bibliography{zhe}

\end{document}